# Non-recurrent Traffic Congestion Detection with a Coupled Scalable Bayesian Robust Tensor Factorization Model ☆

Qin Li[a], Huachun Tan[b], Zhuxi Jiang[c], Yuankai Wu[d], Linhui Ye[e]

[a] School of Mechanical Engineering, Beijing Institute of Technology, Beijing, China
[b] School of Transportation Engineering, Southeast University, Nanjing, China
[c] Faculty of Momenta, Beijing, China
[d] Department of Civil Engineering and Applied Mechanics, McGill University, Montreal, Canada
[e] Civil Engineering in University of Wisconsin-Madison, Madison, Wisconsin, USA

## I. h i g h l i g h t s

· Propose a coupled scalable Bayesian robust tensor factorization model.
· Proposed a new non-recurrent traffic congestion detection method.

## II. a r t i c l e   i n f o

### A B S T R A C T

Non-recurrent traffic congestion (NRTC) usually brings unexpected delays to commuters. Hence, it is critical to accurately detect and recognize the NRTC in a real-time manner. The advancement of road traffic detectors and loop detectors provides researchers with a large-scale multivariable temporal-spatial traffic data, which allows the deep research on NRTC to be conducted. However, it remains a challenging task to construct an analytical framework through which the natural spatial-temporal structural properties of multivariable traffic information can be effectively represented and exploited to better understand and detect NRTC. In this paper, we present a novel analytical training-free framework based on coupled scalable Bayesian robust tensor factorization (Coupled SBRTF). The framework can couple multivariable traffic data including traffic flow, road speed, and occupancy through sharing a similar or the same sparse structure. And, it naturally captures the high-dimensional spatial-temporal structural properties of traffic data by tensor factorization. With its entries revealing the distribution and magnitude of NRTC, the shared sparse structure of the framework compasses sufficiently abundant information about NRTC. While the low-rank part of the framework, expresses the distribution of general expected traffic condition as an auxiliary product. Experimental results on real-world traffic data show that the proposed method outperforms coupled Bayesian robust principal component analysis (coupled BRPCA), the rank sparsity tensor decomposition (RSTD), and standard normal deviates (SND) in detecting NRTC. The proposed method performs even better when only traffic data in weekdays are utilized, and hence can provide more precise estimation of NRTC for daily commuters.

## 1. Introduction

Freeway traffic congestion consists of recurrent congestion, and the additional non-recurrent congestion caused by accidents, breakdowns, and other random events such as inclement weather and debris[1]. Generally, recurrent congestion which is known to many as "rush-hour" traffic is often seen as a capacity problem logically combated with raising roadway capacity and exhibits a daily pattern [2]. The delay caused by recurrent congestions is usually within what commuters expect. However, non-recurrent congestions which occur at arbitrary time of a day due to crashes and incidents, vehicle breakdowns, road construction activities, special events, extreme weather events, bring traffic participants, especially commuters, unexpected delays. Obviously, unexpected delays caused by NRTC make commuters much more frustrated. One explanation is that most travelers are less tolerant of unexpected delays because they cause them to be late for work or important meetings, miss appointments, or incur extra child-care fees. Shippers that face unexpected delay may lose money and disrupt just-in-time delivery and manufacturing processes [3]. Hence, timely knowing and understanding the development of events that lead to non-recurrent congestion plays an important role in helping drivers plan their routes to avoid unexpected delays and improving the management of the transportation infrastructure.

In general, it takes some time for the police department to receive a report and issue a warning to road drivers and traffic managers, and during this time, vehicles may be accumulated due to the events to cause a more severe traffic congestion. Thus, reliable and real-time approaches to detecting or recognizing NRTC based on the observations of traffic data will be desired or required for road drivers and traffic managers to learn NRTC events. Zhang *et al* [4] analyzed and proved the efficiency of loop detectors on traffic data collecting. Hence, the traffic data for detecting NRTC can be observed by loop detectors. The detection results of such approaches are informative for the road drivers to choose routes before falling trapped in the traffic jams. Traffic managers can apply proper managing measures with the result as well.

Many researches have been conducted on the problem of detecting NRTC. Since an incident is usually assumed to cause an NRTC and thus theoretically detection of an incident necessitates detection of an NRTC [5], early research on NRTC detection was usually cast as detecting incidents. Since the mid-1970's, various surveillance-based automotive incident detection algorithms (AID) have been developed [6], but they are vulnerable to severe weather condition. Bayesian [7] and Standard Normal Deviates (SND) algorithms [8][9] were proposed to detect incidents from the statistical aspect based on the relationship between the upstream and downstream traffic conditions. Some researchers used pattern-based algorithms such as the California algorithm [10] and the pattern recognition algorithm [11] by presetting a threshold. In 1994, the fuzzy logic is designed to approximate reasoning when data is missing or incomplete. The fuzzy set logic is applied as a supplement to the California algorithm [12]. In the next year, a time series analysis method ARIMA is introduced into traffic incidents detection, it detects incidents by providing short-term forecasts of traffic occupancies as presetting threshold. With the rise of machine learning, many machine learning methods have also been used to detect traffic incidents, including support vector machine (SVM) [13][14], decision tree [15], integrated algorithm[16][17], and artificial neural networks (ANN)[18-21], almost all of which were supervised leaning methods. For a more comprehensive overview of conventional incident detection algorithms, readers can refer to[22]. In addition, Dr M. Motamed made a detailed description of real-time freeway incident detection models using machine learning techniques in her dissertation [23]. These above-mentioned methods all pursued higher detection rate, lower false alarm rate or less average detection time.

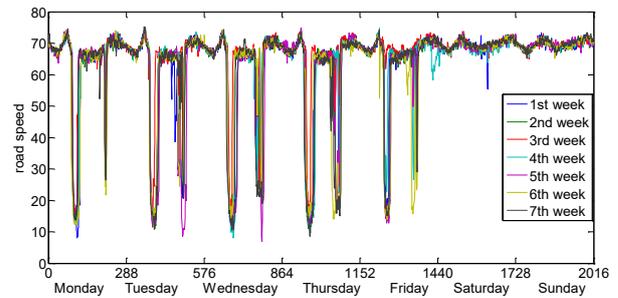

**Fig. 1.** The high-dimensional time-correlation properties of 13-week traffic speed data from Freeway I405

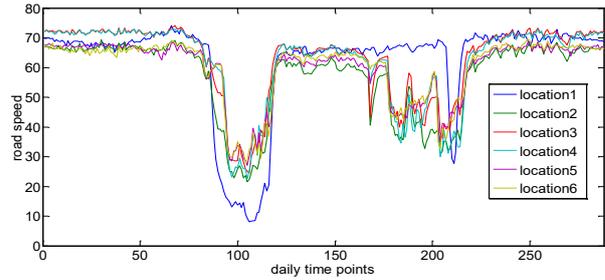

**Fig. 2.** The traffic speed of 6 adjacent detectors. The speed exhibits strong spatial correlations.

However, various factors can cause NRTC and hence detecting NRTC cannot be completely formulated to detecting traffic incidents. Fortunately, all the factors causing NRTC have a similar impact on the traffic condition (data), and thus to effectively detect NRTC, one can focus on traffic condition without the need of considering the factors. The primary data source for quantifying congestion is speed [24], which together with traffic flow and road occupancy can reflect the comprehensive distribution of congestion. Since differences between actual traffic data and normal traffic data indicate the congestion situation caused by non-recurrent events, obtaining the normal traffic data pattern will be a key element for detecting NRTC. Existing presetting-threshold methods employed thresholds (e.g., average values) to define whether traffic data is normal or not at a time interval. When traffic flow or road occupancy is larger than a preset threshold or speed is smaller than a preset threshold at a time, the traffic data is thought of as abnormal. Note that the thresholds vary over time in a day or during a time period, and the pattern of normal traffic data is referred to the distribution of threshold values over the whole-time intervals at a fixed location or over the whole locations at a time interval. Unfortunately, commuters' stationary recurrent commuting behaviors give rise to not only daily pattern containing morning peaks, evening peaks, and off-peak period, but weekly periodic patterns. In addition, traffic data between adjacent locations may be inherently correlated, indicating that traffic data exhibits high-dimensional spatial-correlation properties, as shown in Fig.1 and Fig.2. Fig. 1 depicts the high-dimensional temporal correlation of 13-week traffic speed data, and Fig. 2 depicts The traffic speed of 6 adjacent detectors. The speed shows strong spatial correlations. Consequently, unidimensional patterns defined by simple thresholds are unable to finely express the normal distribution and pattern of high-dimensional traffic data. Machine learning based traffic incidents detection methods learned the traffic incidents through a black box, which cannot give an intuitive comprehend of the normal and abnormal traffic data. Moreover, mainstream supervised machine learning methods require lots of training data to train a detection model, which cannot work when there is no enough historical data. Hence, a training-free model that can fully capture traffic daily, weekly distribution, and spatial correlations is necessitated to better define the complex normal traffic data pattern and to detect non-recurrent congestion without model.

Recently, Yang *et al.* [25] proposed an algorithm detecting road traffic events with the methodology initially designed for detecting moving objects in videos. They continued to present a method of detecting road traffic events by coupling multiple traffic data time series at one location or one type of traffic data at different locations with a nonparametric Bayesian method, Bayesian robust principal component analysis (BRPCA). The method assumes that the distribution of non-recurrent traffic events is sparse in space and time, and the normal distribution of traffic data is low-rank since strong spatial and temporal patterns are proved to be observed on traffic data that imply periodicity and a strong correlation between adjacent upstream and downstream observations. BRPCA proved to be effective in detecting road traffic events based on the traffic flow and occupancy data, measured by 38 loop detector sensors on highway I-494 from the Minnesota Traffic Observatory for the whole year of 2011. However, BRPCA has limitations as follows. 1) It can hardly fully capture traffic data's underlying high-dimensional time-spatial properties simultaneously in the low-rank part. A matrix-based model in BRPCA utilizes only two-modal information and it can either couple two types of traffic data or single type of traffic data from two neighboring detectors, but not both. 2) It employs only occupancy and flow data that are indirectly related to traffic congestion, and thus the detection may not be very accurate since the primary data source for quantifying congestion is speed. 3) Its temporal resolution is 15 minutes (5 minutes are commonly used), which may miss some congestion traffic events, resulting in a higher detection performance than the real case where the temporal resolution is 5 minutes.

In the field of processing unreliable traffic data, Tan. *et al* [26] proved that a tensor model, which has been widely used in face recognition and neuroscience[27], can fully utilize the intrinsic multiple correlations of traffic data, outperforming matrix model based methods. Motivated by this, our previous work proposed a tensor recovery based NRTC detection method [28], which can fully capture underlying high-dimensional time-spatial properties of traffic data simultaneously and hence attained an excellent performance, proving the superiority of tensor recovery model in NRTC detection over matrix-based method RPCA. In[28], we used the tensor recovery method based on minimizing trace norm, the rank sparsity tensor decomposition (RSTD)[29].

However, although having a close-form solution, RSTD relaxed the dependent relationships between each mode into penalty terms, which may result in a solution with lower accuracy. Many tensor recovery methods have also been proposed [31][32]. In particular, Zhao *et al.* [33] recently proposed a variational Bayesian inference and tensor decomposition based tensor recovery method, Bayesian robust tensor factorization (BRTF), which outperforms minimizing trace norm based methods in video background modeling. The problem with RSTD and BRTF lies in that it is difficult for them to couple multiple traffic data in one single tensor model. In order to fully utilize multiple traffic data simultaneously and also employ the tensor model for detecting congestions from the traffic data, this paper proposes a tensor decomposition based tensor recovery model for NRTC detection and recognition. A Gibbs sampling based Scalable Bayesian robust tensor factorization with an automatic rank decrease processing is used through a multiplicative gamma process (MGP). The automatic rank decrease processing and the coupling of multiple traffic data enable the proposed method to provide efficient and accurate NRTC detection and recognition from traffic data.

The rest of the paper is organized as follows. In section 2 the theoretical background is introduced; In Section 3, we analyze the high-dimensional time-spatial properties of traffic data and present the proposed NRTC detection method named Coupled SBRTF; In Section 4, numerical experiment results are given; In Section 5, we discuss the complexity of the proposed method. Conclusions and future work are presented in Section 6.

## 2. Theoretical Background

### 2.1 Tensor basics

A tensor is a multidimensional array. The order of a tensor represents the number of dimensions, also called ways or modes. More formally, an N*th*-order tensor is an element of the tensor product of N vector spaces, each of which has its own coordinate system. Tensors are denoted by boldface Euler script letters, e.g., $\mathcal{X}$. Matrices are denoted by boldface capital letters, e.g., $\mathbf{X}$; vectors are denoted by boldface lowercase letters, e.g., $\mathbf{x}$; scalars are denoted by lowercase letters, e.g., x. Given an N-mode tensor $\mathcal{X} \in R^{I_1 \times I_2 \times \dots \times I_N}$, its $(i_1 \cdots i_n \cdots i_N)th$ entry is denoted as $x_{i_1 \cdots i_n \cdots i_N}$, where $1 \leq i_n \leq I_n, 1 \leq n \leq N$. The "unfold" operation along the *n-th* mode on a tensor $\mathcal{X}$ is defined as:

$$unfold\ (\mathcal{X}, n) := \mathcal{X}_{(n)} \in R^{I_k \times (I_1 \cdots I_{n-1} I_n \cdots I_N)} \quad (1)$$

The Hadamard product denoted by $\odot$, is an elementwise product of two vectors, matrices, or tensors of the same sizes. For example, give matrixes A and B, both of size $I \times J$, their Hadamard product is denoted by $A \odot B$, which is also a matrix of size $I \times J$ as:

$$A \odot B = \begin{bmatrix} a_{11}b_{11} & a_{12}b_{12} & \cdots & a_{1J}b_{1J} \\ a_{21}b_{21} & a_{22}b_{22} & \cdots & a_{2J}b_{2J} \\ \vdots & \vdots & & \vdots \\ a_{I1}b_{I1} & a_{I2}b_{I2} & \cdots & a_{IJ}b_{IJ} \end{bmatrix} \quad (2)$$

### 2.2 CP Tensor Factorization and Bayesian Robust CP Tensor Factorization

The CP (CANDECOMP/PARAFAC) factorization factorizes a tensor into a sum of component rank-one tensors [34]. Given an *N*-order tensor $\mathcal{X} \in R^{I_1 \times I_2 \times \dots \times I_N}$, its CP factorization can be represented by

$$\mathcal{X} = \sum_{r=1}^{R} \lambda_r \cdot u_r^{(1)} \circ u_r^{(2)} \circ \cdots u_r^{(N)} = [\![\lambda; U^{(1)}, U^{(2)}, \cdots U^{(N)}]\!] \quad (3)$$

where $\circ$ denotes the vector outer product, $U^{(n)} = [u_1^{(n)}, u_2^{(n)}, \cdots, u_R^{(n)}]$ is the $n$-th mode factor matrix. Fig.3 gives a CP factorization of a 3-way tensor. With the CP factorization, the tensor element $x_i$ can be represented by

$$\mathcal{X}_i = \sum_{r=1}^{R} (\lambda_r \prod_{n=1}^{N} u_{i_n r}^n) \quad (4)$$

where $\boldsymbol{i} = [i_1, i_2, \cdots, i_n, \cdots, i_N]$ is the N-way index vector. The vector form of the above CP factorization can be represented as

$$\text{vec}(\mathcal{X}) \in R^{n=\prod_{n=1}^{N} i_n} = U^{(1)} \odot U^{(2)} \odot \cdots \odot U^{(N)} \lambda \quad (5)$$

where $\odot$ denotes the Khatri-Rao product and $\lambda = [\lambda_1, \lambda_2, \dots \lambda_R]$ denotes the vector along the super diagonal of the core tensor. For more information about tensor basics, please refer to [33].

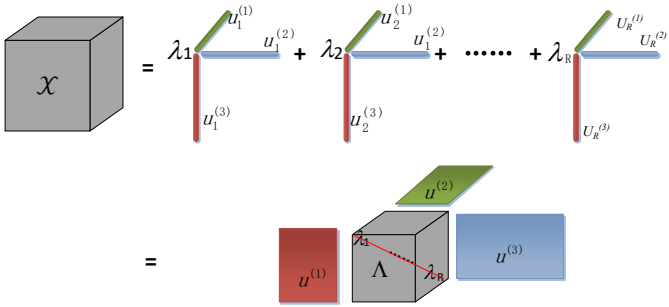

**Fig.3** The CP decomposition for a 3-way tensor with rank R.

Bayesian CP tensor factorization [34-37] is one of the probabilistic tensor factorization methods[38] which factorize a tensor through probabilistic inference. The robust CP factorization is a typical kind of robust tensor factorization represented in Fig. 4, which can factorize a tensor when its data is of both spare noise and dense noise. For example, suppose $\mathcal{Y}$ is an $N$-th order tensor of size $I_1 \times I_2 \times \dots \times I_N$ based on the observed data. Then, it can be represented by the superposition of the true latent tensor $\mathcal{L}$, the sparse outliers $\mathcal{S}$, and the small dense noise $\mathcal{E}$ as: $\mathcal{Y} = \mathcal{L} + \mathcal{S} + \mathcal{E}$, where $\mathcal{L}$ is generated by CP factorization with a low-CP-rank, representing the global information, $\mathcal{S}$ is enforced to be sparse, representing the local information, and $\mathcal{E}$ is usually supposed to be isotropic Gaussian noise. Different from the rank sparsity tensor decomposition (RSTD) algorithm [29], which set the problem to be minimizing the rank of $\mathcal{L}$, the Bayesian robust CP factorization based tensor recovery is to derive the low-CP-rank part $\mathcal{L}$ and the sparse part $\mathcal{S}$ through a probabilistic model by Bayesian inference.

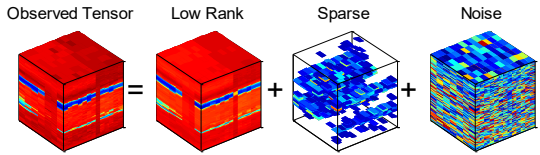

**Fig.4** The framework of robust tensor factorization.

## 3. Coupled SBRTF Based Non-recurrent Traffic Congestion Detection Method

Coupled scalable Bayesian robust tensor factorization based non-recurrent traffic congestion detection method is proposed in this section.

### 3.1 Problem formulation

As mentioned in section II, the NRTC detection and recognition is an important problem for travelers and operators. There are three focal points that need to be noticed in NRTC detection and recognition with a training-free model. First, in [25] and [28], researchers have demonstrated that the NRTC events are similar to the outliers in videos, which are rare and of random sparse property, compared to daily repeated traffic patterns. Hence, the methods of detecting outliers in videos and foreground extraction in images can be used in NRTC detection. Second, according to the traffic flow theory [39], when the NRTC happens for any reasons, multiple types of traffic data including traffic flow, road speed, and road density appear to be abnormal at the same spatiotemporal position, just like a shifting scenery is recorded by several video cameras from different angles and all the videos share the same time and space information of outliers in the scenery. Fig.5 gives a description of one type of traffic pattern of non-recurrent congestion [40], depicting the temporal relationship among traffic flow, speed, and road occupancy when a NRTC happens between time $t_1$ and $t_2$. Therefore, using multiple types of traffic data allows to better capture the characteristics and effects of traffic events which usually cause NRTC. Third, traffic data is temporal-spatial correlated and shows multi-mode features in daily repeated traffic patterns [41].

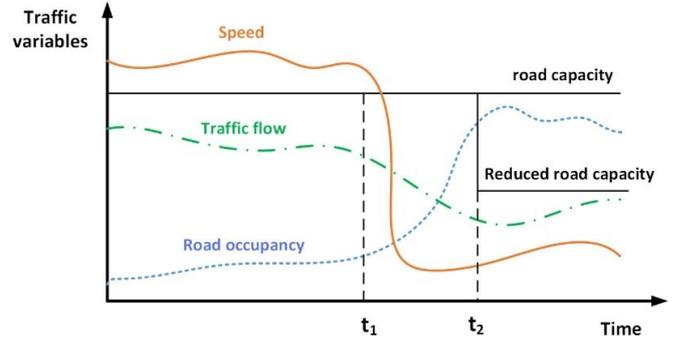

**Fig. 5.** Relationship among traffic flow, speed, and road occupancy at the case of NRTC (happening between time $t_1$ and $t_2$).

The above three points indicate that the first step of detecting NRTC is to define/quantify it based on multiple types of traffic data. The general approach to defining/quantifying NRTC is to define/quantify recurrent congestion on "normal days" (i.e., days that do not have a non-recurring event), and then define/quantify NRTC through comparing traffic condition on days containing non-recurring events with traffic condition on "normal days". The difference between traffic conditions will be the congestion attributable to the non-recurring conditions [24]. Consequently, the distribution of traffic data on "normal days" needs to be precisely modelled for defining NRTC, which contains the recurrent congestion condition and the free-flow condition. For this purpose, the natural temporal-spatial multi-mode properties of repeated normal patterns of multiple traffic data should be fully utilized since otherwise any single-mode properties or single types of traffic data do not encode sufficient information for obtaining the normal distribution.

The BRPCA based method proposed in [25] coupled multiple types of traffic data but only utilized two-mode information of traffic data, which may lead to the inaccurate distribution and hence the accuracy decrease of event detection.

Our previous work [28] proposed that a tensor recovery model utilizing multi-mode properties yielded a high and reliable recognition accuracy in NRTC. However, it did not couple different types of traffic data. Observing the merit and drawback of BPRCA and the previous method, this paper develops a combined method that not only couples multiple types of traffic data but also fully utilizes the multi-mode information of traffic data. The combination allows for the proposed method to detect NRTC fast and accurately. Also, different from what have been researched in [25], this paper directly detects the NRTC rather than various kinds of events causing congestions, since NRTC itself is exactly what travelers are concerned about.

### 3.2 Tensor model for traffic data

We use the traffic flow, road speed, and road occupancy (a surrogate of road density) as the observed variables $\mathcal{Y}_1$, $\mathcal{Y}_2$, $\mathcal{Y}_3$. The three traffic variables can be formulated as four-way tensors $\mathcal{Y}_p \in R^{D \times T \times N \times M}$, p=1, 2, 3, shown in Fig. 6, including ''link mode'', ''week mode'', ''day mode'', and ''interval mode'' [42]. The time interval $T$ is usually set to 5 minutes. Thus, 288 intervals are recorded within a day by each detector. Since there are seven days in one week, we set $D$ to be 7. Then the four-way travel time data based tensor gets to be a tensor as $\mathcal{A}_p \in R^{7 \times 288 \times N \times M}$, $M$ is the number of week preserved and $N$ numbers the links considered.

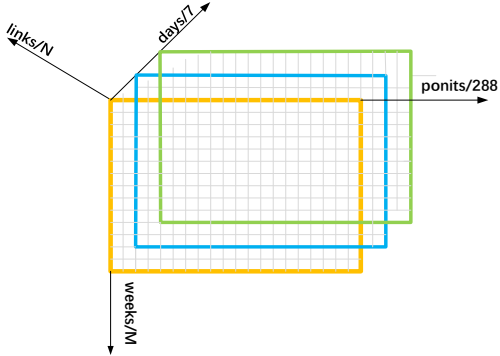

**Fig. 6.** The tensor model for traffic data including the traffic flow, road speed, and road occupancy (a surrogate of road density).

### 3.3 Coupled SBRTF based non-recurrent traffic congestion detection model

Robust tensor factorization based tensor recovery models have been widely used in outliers detecting[28][30]. Given a tensor $\mathcal{Y}$ , a robust tensor factorization model aims to decompose it into a superposition of three components as $\mathcal{Y} = \mathcal{L} + \mathcal{S} + \mathcal{E}$ . The meaning of each component was explained in Section 2. For traffic flow, road speed, and road occupancy, we have

$$\begin{cases} \mathcal{Y}_1 = \mathcal{L}_1 + \mathcal{S}_1 + \mathcal{E}_1 \\ \mathcal{Y}_2 = \mathcal{L}_2 + \mathcal{S}_2 + \mathcal{E}_2 \\ \mathcal{Y}_3 = \mathcal{L}_3 + \mathcal{S}_3 + \mathcal{E}_3 \end{cases} \quad (6)$$

where $\mathcal{Y}_p \in R^{D \times T \times N \times M}$, $p = 1, 2, 3$, represents the observed variables; $\mathcal{L}_p \in R^{D \times T \times N \times M}$, p = 1, 2, 3 , represents the low-rank normally distributed traffic condition; $\mathcal{S}_p \in R^{D \times T \times N \times M}$, p = 1, 2, 3, corresponds to the sparse NRTC, and $\mathcal{E}_p \in R^{D \times T \times N \times M}$, p = 1, 2, 3, represents the dense noise. To extract the same spatiotemporal position where outliers

(NRTC) happens, we replace the sparse part $\mathcal{S}$ with $\mathcal{B} \odot \mathcal{X}$, where $\mathcal{B}$ is a 0-1 tensor describing the distribution of the same sparse spatiotemporal position of the three types of traffic data. Then (6) becomes

$$\begin{cases} \mathcal{Y}_1 = \mathcal{L}_1 + \mathcal{B} \odot \mathcal{X}_1 + \mathcal{E}_1 \\ \mathcal{Y}_2 = \mathcal{L}_2 + \mathcal{B} \odot \mathcal{X}_2 + \mathcal{E}_2 \\ \mathcal{Y}_3 = \mathcal{L}_3 + \mathcal{B} \odot \mathcal{X}_3 + \mathcal{E}_3 \end{cases} \quad (7)$$

where $\mathcal{B} \odot \mathcal{X}_p$, p = 1, 2, 3 , represents the sparse part of $\mathcal{Y}_p$, p = 1, 2, 3. In (7), all the three traffic variables share the same $\mathcal{B}$.

A Bayesian robust tensor factorization model factorizes a tensor from a probabilistic aspect. The model characterizes the information of traffic data vividly through different probabilistic distributions.

Observing this, we use a Bayesian model to derive the latent variables in (7). For each type of traffic data, the low-rank part $\mathcal{L}$ is derived through a scalable Bayesian rank tensor factorization model [43]. Since the CP rank is unknown and the rank estimation for tensors is in general an NP hard problem [44], to obtain the CP rank, instead of expressing the original tensor with a "good-enough" low-rank approximation and finding a convex hull to replace the non-convex NP hard problem[45], e.g., minimizing the trace norm, which may lead to a low accuracy, we infer the CP rank by placing a shrinkage prior, the multiplicative gamma process (MGP) [46] over the super diagonal elements of the core tensor ($\Lambda$ in Fig. 3) in the CP factorization[27].

The MGP shrinkage prior ensures the low-rank property of the each $\mathcal{L}$. In the MGP driven Bayesian CP tensor factorization method (MGP-CP) [43], the MGP prior is represented by

$$\lambda_r \sim \mathcal{N}(0, \tau_r^{-1}), \quad r = 1, \cdots, R$$
$$\tau_r = \prod_{l=1}^{r} \delta_l, \quad \delta_l \sim \Gamma(a_0, 1), \ a_0 > 1 \quad (8)$$

where $\mathcal{N}(\cdot)$ denotes the Gaussian distribution, $\Gamma$ denotes the Gamma distribution. $\tau_r$ increases with $r$, and the precision $\tau_r^{-1}$ of the Gaussian distribution decreases towards 0, which hence makes the mean of $\lambda_r$ goes towards 0 and then ensures the low-rank property of $\mathcal{L}$. For the factor matrix $U^{(n)}$ of the tensor $\mathcal{L}$, we assume that each of its R columns $u_r^{(n)}$ is generated from a Gaussian distribution by

$$u_r^{(n)} \sim \mathcal{N}(u^{(n)}, \Sigma^{(n)}), 1 < r < R; 1 < n < N \quad (9)$$

where $u^{(n)}$ and $\Sigma^{(n)}$ are the mean vector and covariance matrix of the Gaussian distribution of the factor matrix $U^{(n)}$.

For the sparse part $\mathcal{B} \odot \mathcal{X}$ , we use the Bernoulli distribution to model $\mathcal{B}$ as $b_{i_1, i_2, \cdots i_N} \sim \text{Bernoulli}(\pi_{i_1, i_2, \cdots i_N})$, where $b_{i_1, i_2, \cdots i_N}$ is the entry $i = (i_1 \cdots \cdots i_N)$ of $\mathcal{B}$. To ensure its sparsity, we employ the $Beta$ distribution as the conjugate prior of the Bernoulli distribution as $\pi_{i_1, i_2, \cdots i_K} \sim B(\alpha_1, \beta_1)$. $\mathcal{X}$ is supposed to be drawn from the Gaussian distribution by

$$x_{i_1, i_2, \cdots i_K} \sim \mathcal{N}(0, \nu^{-1})$$
$$\nu \sim \Gamma(c_0, d_0) \quad (10)$$

where $c_0$ and $d_0$ is set according to different cases. We assume that the dense noise $\mathcal{E}$ has a Gaussian distribution as

$$e_{i_1, i_2, \cdots i_K} \sim \mathcal{N}(0, \gamma^{-1})$$
$$\gamma \sim \Gamma(e_0, f_0) \quad (11)$$

The goal is to infer the posteriori distribution of all the above parameters and latent variables. There are two typical inference methods: the variational Bayesian (VB) method [33] and the Gibbs sampling method. In general, the Gibbs sampling method provides more accurate inference results than VB, but VB runs faster. Since the MGP-CP ensures the optimization process to converge at a relatively high speed, the Gibbs sampling method is employed in this work. The likelihood of $\mathcal{Y}$ given the other parameters and latent variables can be represented by

$$P(\mathcal{Y}|\{U^{(n)}\}_{n=1}^N, \mathcal{B}, \mathcal{X}, \gamma) = \prod_i \mathcal{N}(\mathcal{Y} | \sum_{r=1}^R (\lambda_r \prod_{n=1}^N u_{in}^{(n)}) + \mathcal{B}_i \odot \mathcal{X}_i, \gamma^{-1}) \quad (12)$$

where $i \in I$ is the entry index. The joint distribution of the model, i.e., $P(\mathcal{Y}, \{\lambda_r\}_{r=1}^R, \{U^{(n)}\}_{n=1}^N, \mathcal{B}, \pi, \mathcal{X}, \gamma, \nu, \tau)$, can be expressed by

$$P(\mathcal{Y}|\{U^{(n)}\}_{n=1}^N, \mathcal{B}, \mathcal{X}, \gamma) \prod_{r=1}^R P(\lambda_r|0, \tau_r^{-1}) \prod_{n=1}^N P(u_r^{(n)}|u_r^{(n)}, \Sigma_r^{(n)}) P(\mathcal{B}|\pi) P(\mathcal{X}|\nu) P(\tau) P(\gamma) P(\pi) p(\nu) \quad (13)$$

We conduct the inference based on the linear Gaussian theorem [47]. Gibbs sampling is used to update the posterior distribution of all model parameters and latent variables given each observation $\mathcal{Y}$ of the three traffic variables. When one variable or parameter is updated, the other variables and parameters are fixed. The parameters and variables are sampled as follows.

The posteriori distribution of $\{\delta_l\} \sim \Gamma(\hat{a}_0, \hat{b}_0)$ :

$$\hat{a}_0 = a_0 + \frac{1}{2}(R - r + 1)$$
$$\hat{b}_0 = b_0 + \frac{1}{2} \sum_{h=r}^R \lambda_h^2 \prod_{h=1, l \neq r}^h \delta_l \quad (14)$$

The posteriori distribution of $\lambda_r \sim \mathcal{N}(\hat{u}_r, \hat{\tau}_r^{-1})$ :

$$\hat{\tau}_r^{-1} = \tau_r + \gamma \sum_i (\prod_{n=1}^N u_{in}^{(n)})^2$$
$$\hat{u}_r = \hat{\tau}_r^{-1} \gamma \sum_i (\prod_{n=1}^N u_{in}^{(n)}) (\mathcal{L}_i - \sum_{r'=1, r' \neq r}^R (\lambda_{r'} \prod_{n=1}^N u_{in}^{(n)})) \quad (15)$$

The posteriori distribution of $u_r^{(n)} \sim \mathcal{N}(\hat{u}_r^{(n)}, \hat{\Sigma}_r^{(n)})$:

$$\begin{cases} \hat{\Sigma}_r^{(n)} = (\Sigma^{(n)-1} + \Phi_r^{(n)})^{-1} \\ \Phi_r^{(n)} = diag(\psi_{1r}^{(n)}, \psi_{2r}^{(n)}, \cdots, \psi_{I_n r}^{(n)}) \\ \psi_{mr}^{(n)} = \gamma \sum_{i, i_n=m} c_{i_n r}^{(n)2}, \quad 1 \leq m \leq I_n \\ \hat{u}_r^{(n)} = \hat{\Sigma}_r^{(n)} (\Sigma^{(n)-1} u^{(n)} + \delta_r^{(n)}) \\ \delta_r^{(n)} = (\delta_{1r}^{(n)}, \delta_{2r}^{(n)}, \cdots, \delta_{I_n r}^{(n)}) \\ \delta_{mr}^{(k)} = \gamma \sum_{i, i_n=m} c_{i_n r}^{(n)} (\mathcal{L}_i - d_{i_n r}^{(n)}), \quad 1 \leq m \leq I_n \end{cases} \quad (16)$$

In (14), (15), and (16), $1 \leq r \leq R$, $1 \leq n \leq N$, and $i=(i_1, i_2, \cdots i_N)$.

The posteriori distribution of $\mathcal{B}_i \sim Bernulli(q_1/(q_1 + q_0))$:

$$q_1 = \pi_i \exp(-0.5\gamma(\mathcal{X}_i^2 - 2\mathcal{X}_i \delta_i))$$
$$q_0 = \pi_i \exp(-0.5\gamma(\delta^2)) \quad (17)$$

The posteriori distribution of $\pi_i \sim \Gamma(\hat{\alpha}_1, \hat{\beta}_1)$ :

$$\hat{\alpha}_1 = \alpha_1 + \mathcal{B}_i$$
$$\hat{\beta}_1 = \beta_1 + 1 - \mathcal{B}_i \quad (18)$$

The posteriori distribution of $\mathcal{X}_i \sim \mathcal{N}(u, \Sigma)$:

$$\Sigma \leftarrow (\nu + \gamma \mathcal{B}_i^2)^{-1}$$
$$u \leftarrow \Sigma \mathcal{B}_i^T \gamma \delta_i \quad (19)$$

The posteriori distribution of $\nu \sim \Gamma(\hat{c}_0, \hat{d}_0)$:

$$\hat{c}_0 \leftarrow c_0 + 0.5 I_1 \cdots I_N$$
$$\hat{d}_0 \leftarrow d_0 + 0.5 \sum_i \mathcal{X}_i^2 \quad (20)$$

The posteriori distribution of $\gamma \sim \Gamma(\hat{e}_0, \hat{f}_0)$:

$$\hat{e}_0 \leftarrow e_0 + 0.5 I_1 \cdot I_2 \cdots I_N$$
$$\hat{f}_0 \leftarrow f_0 + 0.5 \sum_i (\mathcal{Y}_i - \mathcal{L}_i - \mathcal{B}_i \circ \mathcal{X}_i)^2 \quad (21)$$

In (17)-(21), $1 \leq n \leq N$ and $i=(i_1, i_2, \cdots i_N)$.

---

**Algorithm 1:** Coupled Scalable Bayesian Robust Tensor Factorization

**Input:** $p$ numbers of Nth-order tensor $\mathcal{Y}_p$

**Initialization:** $\{u_r^{(n)}, \Sigma_r^{(n)}, \forall n \in [1, N], \mathcal{B}, \mathcal{X}, \mathcal{E}, \lambda, R, \varphi, \beta_0, \beta_2$; hyperparameters $\nu, \pi, \gamma, \tau$; top level hyperparameters $a_0, b_0, c_0, d_0, e_0, f_0, \alpha_0, \alpha_1, \beta_1, \delta_r, \forall r \in [1, R]$ **for** each $\mathcal{Y}_p, \forall p \in [1, P]$.

  **for** iter = 1 to $N_{Burn-in} + N_{Collec}$

  **for** each $\mathcal{Y}_p$
    **for** r=1 to R
     **for** n=1 to N **do**

      Update the posterior of $u_r^{(n)}$ with (16);
     **end for**
    Update the posterior of $\delta_r$ by (14);
    Update the posterior of $\lambda_r$ by (15);
    Update the posteriori distribution of $\mathcal{B}_i$ by (17);
    Update the posteriori distribution of $\pi_i$ by (18);
    Update the posteriori distribution of $\mathcal{X}_i$ by (19);
    Update the posteriori distribution of $c_0, d_0, e_0, f_0$ by (20-21);
    **if** iter $> N_{Collec}$, collecting, **end**.
  **end**

In the sampling steps, $i=(i_1, i_2, \cdots i_N)$ is the multiple index, and $\mathcal{L}_i$, $\mathcal{B}_i$, $\mathcal{X}_i$ and $\pi_i$ is the $ith$ entry of $\mathcal{L}$, $\mathcal{B}$, $\mathcal{X}$ and $\pi$, respectively. An adaptive process is employed in the MGP - the component tensor with $|\lambda_r| < \varphi$, $1 \leq r \leq R$, is removed from the model if $\lambda_r$ becomes smaller than a predefined threshold $\varphi$, otherwise a new component tensor will be added if $|\lambda_r| > \varphi$, for any $r \in [1, R]$. Such an adaptive process occurs in a probability $p(t) = e^{(\beta_0 + \beta_2 t)}$ at the $t^{th}$ iteration, such that $p(t)$ is approximately 0.1 when $t$ is relatively small and then decreases exponentially with $t$. Denote by $N_{Burn-in}$ and $N_{Collec}$ The number of burn-in iterations and the number of colleting samplers are denote by $N_{Burn-in}$ and $N_{Collec}$, and the whole procedure of model inference is summarized in Algorithm 1.

In the proposed Couple SBRTF based non-recurrent traffic detection congestion model, three types of traffic data are coupled. To understand the coupling, Fig. 7 illustrates a simple process in which only two types of traffic data are coupled, p=1, 2, (e.g., traffic flow and speed).

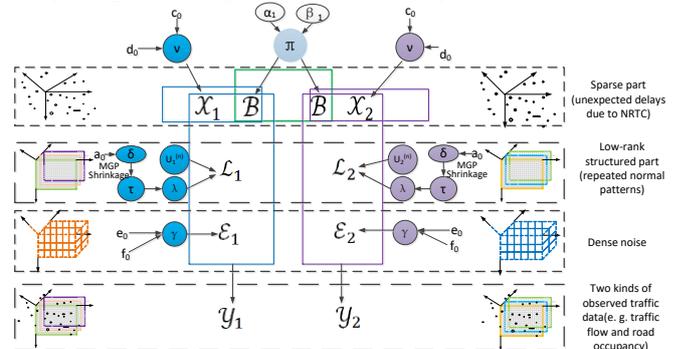

**Fig. 7.** The Coupled SBRTF based non-recurrent traffic congestion detection model for the case in which two different types of traffic data are coupled, p=1, 2, (e.g. traffic flow and speed).

## 4. Numeral Experiments

In this section, numeral experiments are conducted to verify the feasibility of the proposed method.

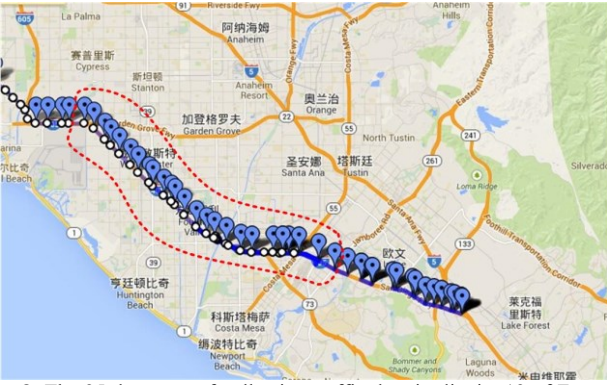

**Fig. 8.** The 25 detectors of collecting traffic data in district 12 of Freeway I405.

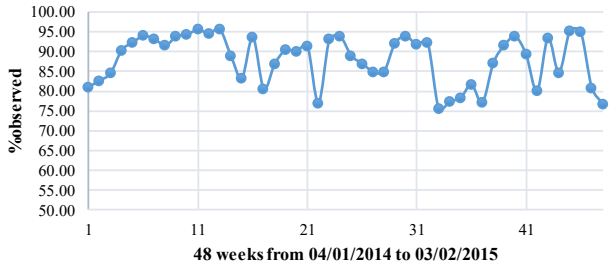

**Fig. 9.** The 5-min observation of lane points on the 14-mile long segment of the North-bound I-405 trip

### 4.1 Test database

We use the raw traffic flow, road speed, road occupancy data and traffic incident reports from April 1st, 2014 to March 2nd, 2015 attained from the Caltrans PeMS (http://pems.dot.ca.gov/) database. PeMS provides a consolidated database of traffic data collected by Caltrans placed on state highways throughout California, as well as other Caltrans and partner agency data sets. As for the incident reports, PeMS provides California Highway Patrol (CHP) computer-aided dispatch (CAD) incident reports as well as Traffic Accident and Surveillance Analysis System (TASAS) data reports. The CHP incident reports include all incident found in the CHP CAD. The TASAS records include all accidents that occur on State Highways. The TASAS records are manually verified by Caltrans staff. For more details, please refer to the PeMS User Manual [48]. Three types of traffic data and the traffic incident (CHP and TASAS) reports were collected from 25 locations (framed by the red dotted line) in an about 14-mile long segment of the North-bound I-405 trip, as shown in Fig. 8. The daily average distance between adjacent detectors with observed data varies from 0.17 to 1.43 miles. The original quality of the raw traffic data utilized (% observed) in this paper provided by the PeMS is showed Fig. 9, from which we can see that the percent of observed data is no less than 75% on each week. Meanwhile, we searched for the observation of each detector at each time point, there is no "failure" detector with its percent of observed data being 0%. As for the missing data, PeMS made a simple imputation using the local neighbors, the global neighbors, the cluster medians or the temporal medians. The traffic events recorded in the traffic events reports serve as the ground-truth data.

The ground-truth events include traffic accidents, hazards, breakdowns, police control, congestion, and bad weather.

### 4.2 Experimental Results for One or Two Detectors on All Days

In this paper, we compare the proposed method with three methods. The first method is a traditional well-known non-recurring congestion measuring method called *Standard Normal Deviate Algorithm* (SND) [1]. The second one is *Bayesian Robust Principle Component Analysis* (BRPCA) [24], which has been used to detect road traffic incidents in 2014. The third one is a tensor recovery method called *Rank Sparsity Tensor Decomposition* (RSTD) [29], which has been widely used to detect outliers. Another reason to compare with RSTD is that RSTD is a very classical tensor recovery method and it has close-form solutions (analytic solutions), and comparison based analysis make sense if the analytic solutions of the problem are available.

For each method, we count the number of the detected events, the detected but unlabeled events, and the undetected but labeled events (false positives). It's worth noting that, there are two types of traffic events detected in the original paper of the method BRPCA, however, since we only care about the non-recurrent congestion here which corresponds to the type one, we ignore the other type.

For BRPCA, the data of one or two detectors are used for testing the detection performance. Therefore, to properly compare the proposed method with BRPCA, RSTD, and SND, this paper utilizes one or two of the 25 detectors for the experiments in Section 4-B and Section 4-C. But SND only works for the data of one detector, hence for case of using two detectors, we only compare the proposed method with the coupled BRPCA and RSTD.

For the case of one detector, the 11th detector is used, on which there are totally 79 recorded traffic events, each causing traffic congestion over 4 minutes. Both the proposed method and the Coupled BPRCA couple traffic flow, road occupancy, and speed data. RSTD and SND use traffic data separately, i.e., they use only one of traffic flow, speed, and road occupancy data at a time. The proposed Coupled SBRTF method constructs the whole traffic data of the detector as three $288 \times 7 \times 48$ sized tensors. The initial rank of the three tensors is randomly set to 15. The hyperparameter $a_0$ for the multiplicative gamma process is set to $[4.5, 3, 1.5]$, and all the initial values of the other hyperparameters are set to $10^{-6}$, except that the shape parameter and dimension parameter of $\pi_i$ are set to 0.001 and 0.999. $N_{\text{Burn-in}}$ and $N_{\text{Collec}}$ are set to 1500 and 500. The Coupled BRPCA method constructs the whole traffic data of the detector as three $288 \times 336$ sized matrices. The initial rank of the three matrices is set to the largest possible rank 288. The value of $N_{\text{Burn-in}}$, $N_{\text{Collec}}$, and hypermeters are as the same as that of the Coupled SBRTF method. RSTD constructs the traffic flow, speed or road occupancy of the detector as a $288 \times 7 \times 48$ sized tensor. The parameters of RSTD are set to achieve the optimum efficiency as mentioned in the original paper, with the maximal number of iterations being 2000 and the tolerance being $10^{-6}$. SND constructs the traffic flow, speed or road occupancy of the detector as a $2016 \times 48$ sized matrix. The standard deviation is set to 1.5 such that SND find roughly the same number of false-positive non-recurrent congestion events as the other methods.

**Table 1** Detection Accuracy Comparison of The Four Methods Using Data from The Same One Loop Detector

| Methods | Variable | Ratio |
|---|---|---|
| SND | speed | 69.62% |
| SND | occupancy | 65.82% |
| SND | flow | 64.56% |
| RSTD | speed | 81.01% |
| RSTD | occupancy | 82.28% |
| RSTD | flow | 79.75% |
| Coupled BPRCA | flow, occupancy, speed | 83.54% |
| Coupled SBRTF | flow, occupancy, speed | 86.08% |

**Table 2** Detection Accuracy Comparison of The Four Methods Using Data from The Same Two Neighboring Loop Detectors

| Methods | Variable | Ratio |
|---|---|---|
| Coupled BPRCA | speed | 76.92% |
| Coupled BPRCA | occupancy | 75.00% |
| Coupled BPRCA | flow | 72.11% |
| RSTD | speed | 82.09% |
| RSTD | occupancy | 81.73% |
| RSTD | flow | 78.84% |
| Coupled SBRTF | flow, occupancy, speed | 84.61% |

Table 1 shows the accuracy, defined to be the ratio of the number of detected congestion events over the ground-truth events for the four methods. It can be seen from Table 1 that the proposed method outperforms the Coupled BPRCA, RSTD, and SND. It provides the highest accuracy of detecting NRTC, 86.08%, followed by the Coupled BPRCA that gives 83.54%, while RSTD with occupancy gives 82.28% and SND with traffic flow provides 64.56%.

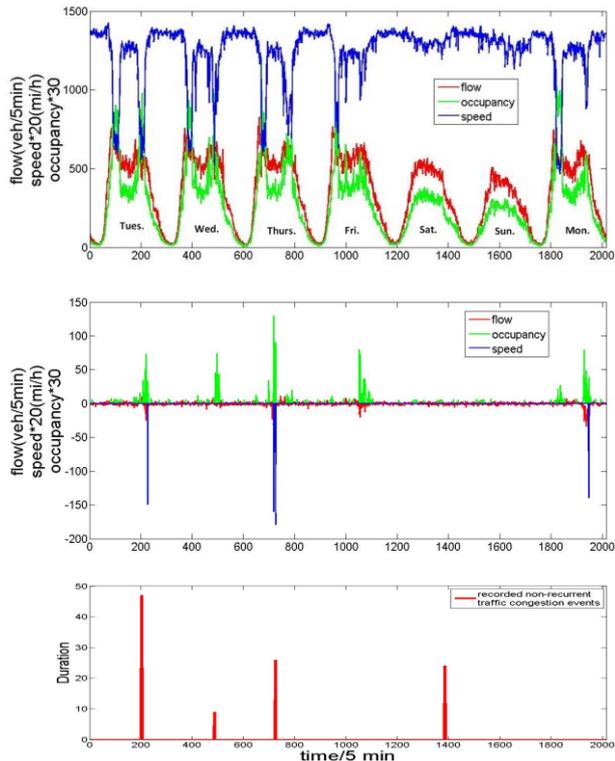

Fig. 10. The detecting results of Coupled SBRTF for 1week with all days on the *9th* detector. Top row represents the raw traffic data, the second row represents the sparse part (the detected non-recurrent congestion), the third row represents the recorded NRTC events.

For the case of two neighboring detectors, the 8[th] and the 9[th] detectors are used, on which there are totally 104 recorded traffic events, each causing traffic congestion over 4 minutes. The Coupled SBRTF constructs the whole traffic data of the two detectors as three $288 \times 7 \times 48 \times 2$ sized tensors. The Coupled BPRCA and RSTD use traffic data separately, i.e., they use only one of traffic flow, speed, and road occupancy data at a time. The Coupled BPRCA constructs the traffic flow, speed or road occupancy of the two detectors as two $288 \times 336$ sized matrixes. RSTD constructs the traffic flow, speed or road occupancy of the two detectors as a $288 \times 7 \times 48 \times 2$ sized tensor. The parameters and hyperparameters are set in a similar manner to the case for one detector.

Table 2 shows the accuracy for the case of two detectors. The proposed method provides the highest accuracy of detecting NRTC events, 84.61%, followed by the RSTD with speed data that gives 82.09%. The number of false positive congestions of the Coupled BRPCA (with speed data), RSTD (with speed data), and the proposed Coupled SBRTF were 98, 93 and 86, respectively.

### 4.3 Experimental results for one or two detectors on only weekdays

Through analysis, we find that, for the RSTD, Coupled BRPCA, and Coupled SBRTF methods, many undetected NRTCs happen on weekends, especially for the RSTD and the proposed method. For example, Fig.10 shows the detecting results of Coupled SBRTF for 1week with all days on the *9th* detector, in which an NRTC happening on Saturday was not detected, but all the other NRTCs have been correctly detected. It is noting that, to give a more intuitive depiction of three types of traffic data in one figure, we multiply speed by 20, and occupancy by 30. The reason might be that the relevancy of traffic condition among weekdays is stronger than the relationship between weekdays and weekends due to commuting requirements. Hence, in this part, we conduct experiments that compare the four methods only on weekdays. Similar to the experiments on all the days, the experiments for one detector and for two neighboring detectors are conducted separately. In the case of one detector, the number of total recorded traffic events is 67. The Coupled SBRTF method constructs the whole traffic data of the 11th detectors as three $288 \times 5 \times 48 \times 2$ sized tensors. The Coupled BRPCA method constructs the three kinds of traffic data as three $288 \times 240$ sized matrixes. The RSTD constructs the speed or road occupancy data of this detector as a $288 \times 5 \times 48 \times 2$ sized tensor. The SND method constructs the traffic flow, speed or road occupancy data as a $288 \times 240$ sized matrix. The number of false positive congestions of SND (with occupancy data), the Coupled BRPCA (with traffic flow, road occupancy, and speed data), RSTD (with occupancy data), and the proposed Coupled SBRTF were 91, 96, 90, and 91, respectively.

The parameters and the experimental settings for the case of two detectors on weekdays are similar to the case of two detectors on all the days. The number of total recorded traffic events is 96. The number of false positive congestions of the Coupled BRPCA (with speed data), RSTD (with speed data), and the proposed Coupled SBRTF were 83, 78, 72, respectively.

**Table 3** Detection Accuracy Comparison of The Four Methods Using Data on Only Weekdays from The Same One Loop Detector

| Methods | Variable | Ratio |
|---|---|---|
| SND | speed | 52.24% |
| SND | occupancy | 71.64% |
| SND | flow | 49.25% |
| RSTD | speed | 85.07% |
| RSTD | occupancy | 86.58% |
| RSTD | flow | 83.87% |
| Coupled BPRCA | flow, occupancy, speed | 85.07% |
| Coupled SBRTF | flow, occupancy, speed | 91.04% |

**Table 4** Detection Accuracy Comparison of The Four Methods Using Data on Only Weekdays from The Same Two Neighboring Loop Detectors

| Methods | Variable | Ratio |
|---|---|---|
| Coupled BPRCA | speed | 79.57% |
| Coupled BPRCA | occupancy | 77.42% |
| Coupled BPRCA | flow | 72.04% |
| RSTD | speed | 84.94% |
| RSTD | occupancy | 83.87% |
| RSTD | flow | 81.72% |
| Coupled SBRTF | flow, occupancy, speed | 87.09% |

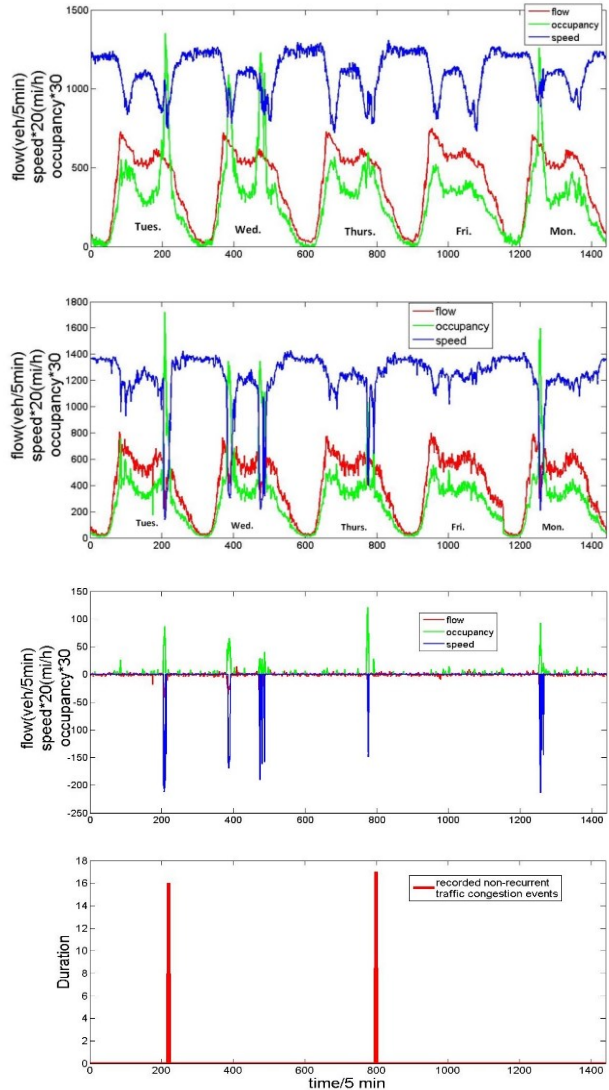

**Fig. 11.** The raw traffic condition, the expected traffic condition, the detected NRTC, and the recorded NRTC on weekdays of 1week of the 11th detector extract by Coupled SBRTF. Top row represents the raw traffic data, the second row represents the low-rank part (the expected traffic condition), the third row represents the sparse part (the detected non-recurrent congestion), the fourth row represents the recorded NRTC events.

Table 3 shows the accuracy for one detector on weekdays. Fig.11 gives detecting details containing the raw traffic condition, the expected traffic condition, the detected NRTC, and the recorded NRTC on weekdays of 1week at the 11th detector. Table 4 shows the accuracy for two neighboring detectors on weekdays. It can be seen from Table 3 and Table 4, the proposed method outperforms SND, RSTD and the Coupled BRPCA in the case of one detector. It gives the highest accuracy 91.04%, followed by RSTD with occupancy that provides 86.58%, and the Coupled BPRCA gives 85.07%, while SND with traffic flow still gives the lowest accuracy 49.25%, which is even inferior to the accuracy it provided in the case of all days. In the case of two detectors,

Coupled SBRTF provides 87.09%, RSTD with speed gives 84.94%, and the Coupled BRPCA with speed gives 79.57%. It is interesting to find, for RSTD, utilizing occupancy data provide higher accuracy than utilizing traffic flow or speed data in the case of one detector, while it gives the best accuracy with speed data in the case of two detectors. From Fig.11, we can see, we detected slight traffic flow slow down, sharp speed slow down, and sharp road occupancy uprush at the same time-spatial position with the two recorded NRTC events. Besides, we also detected three other non-recurrent traffic congestion events that have not been recorded but lead outliers to the raw data and changes the traffic condition. For the detected but not recorded NRTC events, we infer that, there might be a congestion that was omitted by the operator, or the event only lead to slight speed slowdown, which did not cause a congestion duration.

### 4.4 Experimental results for all 25 detectors on all days and only weekdays

Since the proposed method is the only method which can both couple multiple traffic data and meanwhile capture the high-dimensional multi-modal properties of traffic data, we use all the traffic data from the 25 detectors to detect the whole 427 events. The results show that, there are 364 events are detected. The detecting accuracy is 85.23%, and the number of false positive congestions is 344. And when only weekdays are considered, the totally detecting accuracy is 88.43%, and the number of false positive congestions is 291. Fig.12 gives a total detecting results of our proposed method, from which we can see that when multiple neighboring detectors are used, the accuracy improves a little but not so much than only considering two detectors.

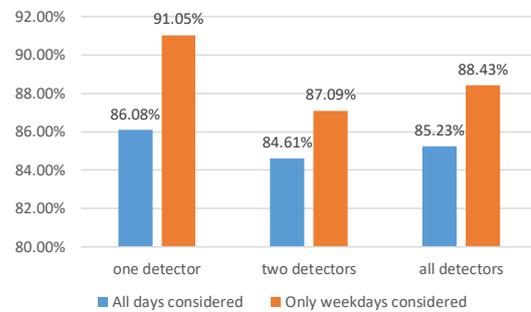

**Fig.12.** Detecting accuracy of our method in this paper in different cases

### 4.5 The influence of system uncertainties (system noises)

It is inevitable that traffic data from detectors may be disturbed by the environmental noises which may cause internal noises to the model. Hence, we add a noise term (i.e. $\mathcal{E}$) in our model, which is supposed to be independent identically distributional (i.i.d.) and zero-mean. In [49], O. Tutsoy, and S. Colak proposed an adaptive estimator design for unstable output error systems, which performs pretty well. They pointed out that it is meaningful to make a test on the system robustness under different uncertainties for a non-parametric model. To make a validation on the model robustness and efficiency under various kinds of uncertainties, we conduct a simple test on traffic data added with extra Gaussian distributional noises. The Gaussian distributional noise is draw from $\mathcal{N}(0, 0.01)$ (case 1) or $\mathcal{N}(0,$

0.02) (case 2). That means for each kind of observed traffic data $\mathcal{Y}$, we have:

$$\begin{cases} \mathcal{Y} = \mathcal{Y} + \sqrt{0.01 \cdot \mathrm{var}(\mathrm{vec}(\mathcal{Y}))} \cdot \mathrm{randn}(\mathrm{DIM}) & case\ 1 \\ \mathcal{Y} = \mathcal{Y} + \sqrt{0.02 \cdot \mathrm{var}(\mathrm{vec}(\mathcal{Y}))} \cdot \mathrm{randn}(\mathrm{DIM}) & case\ 2 \end{cases}$$
(22)

where $\mathrm{var}(\cdot)$ stands for the variance, $\mathrm{vec}(\cdot)$ denotes the vectorization operator, $\mathrm{randn}(\cdot)$ is the standard normal distribution, and DIM is the size of $\mathcal{Y}$. Since SND does not concern about the noise, we only make comparison on the proposed Coupled SBRTF, RSTD, and the Coupled BRPCA. We choose the worst condition for example, i.e. using data from the same two loop detectors. For RSTD and the Coupled BRPCA, the speed data is used, which makes them perform better. Fig.13 gives the results of different methods under the three cases of system uncertainties, where case 0 means using original data with no extra noise.  It can be seen that the three methods are all relatively robust to the small system Gaussian noise, especially RSTD and the proposed coupled SBRTF. The results difference between case 1 and case 2 are much smaller than that between case 0 and case 1.

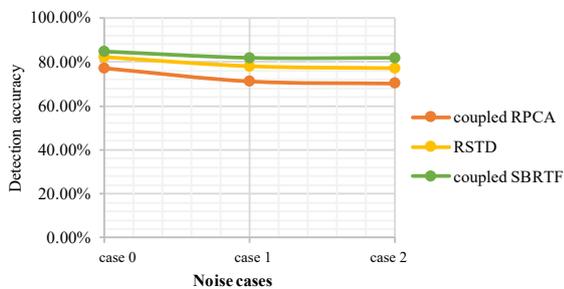

**Fig.13.** Detecting accuracy of different methods under small system Gaussian noises

### 4.6 Total analysis of experimental results

The experimental results show that, from the point of detection accuracy of recorded traffic events, the proposed Coupled SBRTF method outperforms other methods in most occasions, especially in the case of considering only weekdays, and the RSTD method also attains relative higher accuracy on only weekdays. We guess that, it is because there exists strong similarity between the traffic data in the daily mode on weekdays. All the methods except SND perform better in the case of using one detector than in the case of using the two detectors. This is probably because that there exist off-ramps between the two detectors, which can affect the correlations between the two detectors and hence influence the NRTC detecting accuracy for the methods that have utilized the spatial correlations. This phenomenon has been improved by the proposed coupled SBRTF using all the detectors. The coupled SBRTF gives a moderate performance in the case of using all the detectors, which is better than the performances given by all the methods in the case of using two detectors and the performances gotten by the other methods in the case of using one detector. However, the performance of coupled SBRTF using all the detectors is still not as good as that of coupled SBRTF using one detector. The reasons could be that there are many long-duration congestions in that detector which are more easy to be detected, and there may exist on-ramps and off-ramps between some other adjacent detectors.

As for the false positive congestions, from Table 5, we can see that tensor based methods (i.e. RSTD and Coupled SBRTF) outperform the matrix based method (Coupled BRPCA). We set the pre-thresholds for SND so that it detected approximately the same number of false positive NRTC events with the other methods. In the case of one detector, RSTD detected a little smaller number of false positive congestions than Coupled SBRTF. In the case of two neighboring detectors, the proposed method in this paper performs the best.

The validation experimental results show that the three methods are all relatively robust to the small system Gaussian noise, especially RSTD and the proposed coupled SBRTF.

Actually, there are some recorded traffic incidents that did not impose severe impacts on the traffic data, and hence did not lead to NRTC, and meanwhile there are some NRTC events that have not been recorded in the logs. But we have no way to check and verify the real conditions, and just deduce that if all these are taken into account, a higher detection accuracy may be attained.

**Table 5** False   Positive Numbers of Different Methods in Different Cases

|  | One detector | | Two neighboring detectors | |
|---|---|---|---|---|
| **All days** | Recorded events | 79 | Recorded events | 104 |
| | SND (occupancy) | 112 | SND | ---- |
| | Coupled BPRCA | 108 | Coupled BPRCA (speed) | 98 |
| | RSTD (occupancy) | 103 | RSTD | 93 |
| | Coupled SBRTF | 106 | Coupled SBRTF | 86 |
| **Only Weekdays** | Recorded events | 67 | Recorded events | 96 |
| | SND (occupancy) | 91 | SND | ---- |
| | Coupled BPRCA | 96 | Coupled BPRCA (speed) | 83 |
| | RSTD (occupancy) | 90 | RSTD (speed) | 78 |
| | Coupled SBRTF | 91 | Coupled SBRTF | 72 |

## 5. Discussion

### 5.1 Computation efficiency of the coupled SBRTF approach

One of the advantages of the proposed Coupled SBRTF method is that it employs the multiplicative gamma process (MGP) in the low-rank part, in which $\lambda_r$ shrinks to zero with the increasing number of columns in the matrix factors, and hence ensures the low-rank properties. And an adaptive process is employed in the MGP, so we do not have to impose a large enough rank to the initial rank like what has been set for the BRPCA and some other Gibbs sampling based tensor recovery methods. Such an adaptive process can improve the computing speed and decrease the time cost of one iteration. The experiments were running on a 64-bit machine with 16-GB memory and MATLAB 2013b environment. The per-iteration computational cost of the proposed algorithm testing on one about year of traffic data is about 0.36 seconds. Besides, the per-iteration computational cost of the proposed algorithm is scalable, which is linear in the number of observation. And in practical NRTC congestion, only a few weeks or several time intervals of traffic data need to be used, the computational cost would be much less. Hence, the proposed method is efficient and scalable in time cost at NRTC detecting in reality.

## 5.2 Online non-recurrent traffic congestion detection

Since the computational cost is acceptable, we can consider online NRTC detection by the proposed method. We have ever proposed an online travel time prediction method and an online traffic flow prediction method based on dynamic tensor completion [50][51]. The flame of the incremental dynamic tensor analysis can also be used to realize online NRTC detecting. In fact, there are many other dynamic flames, which can be lucubrated in the future research work.

## 6. Conclusion

In this paper, we proposed a training-free novel non-recurrent traffic congestion detection method, namely Coupled Scalable Bayesian Robust Tensor Factorization (Coupled SBRTF). The proposed Coupled SBRTF fully captures the underlying high-dimensional time-spatial properties of traffic data through a tensor model, and couples multiple traffic data simultaneously through a sparsity-sharing structure. Besides, the proposed method employs a multiplicative gamma process (MGP) in the low-rank part, which ensures the low-rank properties of the normal traffic pattern and provides a higher computation efficiency than the conventional Gibbs sampling. The multiplicative gamma process (MGP) had been proposed and used in tensor completion and matrix completion before, however, it is the first time that it is employed in the Robust Tensor Factorization, combined with the actual practice demand.

Experiments on real traffic datasets show that Coupled SBRTF generally outperforms the conventional and the state-of-art dynamic training-free NRTC detection methods including SND, Coupled BRPCA, and RSTD, although RSTD detected a little smaller number of false positive congestions than Coupled SBRTF in the case of using one detector. The proposed method performs even better when only traffic data in weekdays are utilized, and hence can provide more precise estimation of NRTC for daily commuters. Besides, the proposed method extracts the distribution of general expected traffic condition, containing traffic flow, speed and occupancy, as an auxiliary product for commuters and researchers through the low-rank structure, which has not been achieved in existing works. It is interesting to find that RSTD utilizing occupancy data provides higher accuracy than it utilizes traffic flow or speed data in the case of one detector, while it gives the best accuracy with speed data in the case of two detectors. Another funny discovery is the reasonable robustness to the small system noise of the three methods which considered system noises in the models, especially RSTD and the coupled SBRTF.

Through the experimental results, we find that, the existence of off-ramps and on-ramps may affect the distribution of traffic data, which leads to impacts on the accuracy of detection, hence, in the future work, we will take into account the traffic data information of the off-ramps and on-ramps. Besides, we will consider different probability distribution of the shared sparse structure, e.g. polynomial distribution, to define and recognize the level of congestion in detail.

## 7. Acknowledgments

The research was supported by NSFC (Grant No. 61620106002, 51308115). And the authors wish to express their appreciation and gratitude to their colleague, Yong Li (an associate professor working with Beijing University of Posts and Telecommunications, China), for reading the manuscript and his suggestions.

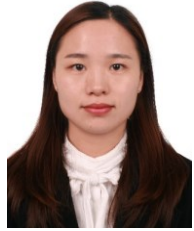

**Qin Li** received the Bachelor's degree from the Department of Transportation Engineering, Beijing Institute of Technology, Beijing, China, in 2014. She is pursuing the Ph.D. degree with the School of Mechanical Engineering, Beijing Institute of Technology, and was a visiting student with Department of Civil & Environmental Engineering, University of Wisconsin-Madison from Oct. 2017 to Oct, 2018. Her research interests include machine learning and intelligent transportation systems.

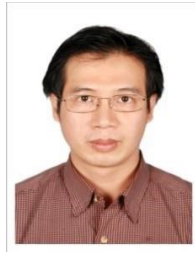

**Huachun Tan** (M'13) received the Ph.D. degree in electrical engineering from Tsinghua University, Beijing, China, in 2006. He used to be an Associate Professor with the School of Mechanical Engineering, Beijing Institute of Technology, Beijing from Sep. 2009 to June 2018. He is now a Professor with the School of Transportation, Southeast University, Nanjing, China. His research interests include image engineering, pattern recognition, and intelligent transportation systems.

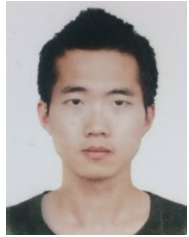

**Zhuxi Jiang** received the Master's degree from the Department of Transportation Engineering, Beijing Institute of Technology, Beijing, China, in 2018. He is currently an algorithm engineer in Momenta. His research interests include machine learning, computer vision, and intelligent transportation systems.

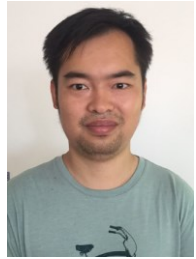

**Yuankai Wu** received the PhD's degree from the School of Mechanical Engineering, Beijing Institute of Technology, Beijing, China, in 2019. He was a visit PhD student with Department of Civil & Environmental Engineering, University of Wisconsin-Madison from Nov. 2016 to Nov, 2017. He is going to be a Postdoc researcher with Department of Civil Engineering and Applied Mechanics of McGill University. His research interests include intelligent transportation systems and machine learning.

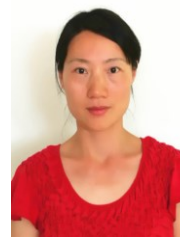

**Linhui Ye** received the M.S. degree in Information and Telecommunication Engineering from the Beijing University of Posts and Telecommunications, Beijing, China, in 2007. She worked in IBM from 2007 to 2017 focusing on software development and testing. She is currently pursuing the Ph.D. degree with the Civil Engineering in University of Wisconsin-Madison, Madison, WI, USA. Her research interests include intelligent transportation systems and machine learning.